\documentstyle[12pt,epsf]{article}
\title{Isospin Dependence of Proton and Neutron Radii
           within Relativistic Mean Field Theory
\footnote{This work is partly supported by the Polish Committee
          of Scientific Research under Contract No. 2~P03B 049 09.}}

\author{           M. Warda, B. Nerlo-Pomorska and K. Pomorski\\
   Theoretical Physics Department, University M.C.S., Lublin, Poland}
\date{}

\begin{document}

\maketitle

\begin{abstract}
The binding energies, shapes and sizes of even-even $\beta$--stable nuclei
with $A\geq 40$ and a~few chains of isotopes with $Z$=50, 56,
82, 94 protons and isotones with $N$=50, 82, 126 neutrons are
analyzed. The average isospin dependence of the radii of protons
and neutrons evaluated within the relativistic mean field theory
is studied.  A simple, phenomenological formula for neutron
radii is proposed.\\


\end{abstract}

\section{Introduction}

The nuclear radii are ones of the crucial quantities good for testing
of every theoretical model of nucleus. The radii are measured with a high
accuracy for the charge density distributions \cite{ott} and less precise
for the neutron ones \cite{batt}. The data concerning nuclear sizes
and shapes are already known for many nuclei, especially for those close to
the $\beta$ stability line, but also more exotic nuclei are
explored extensively by experimentalists at present. It would be
worthwhile to know what the relativistic mean field  theory (RMFT)
predicts for such nuclei  and to approximate the results in terms of
a simple, practical formulae for radii.

The nuclear radii depend chiefly on the number of nucleons ($A$) \cite{boh}
\begin{equation}
   R_0 = r_0 A^{1/3}\,\,,
%1
\end{equation}
where the radius constant $r_0\approx$1.2 fm.  It is the result
of saturation property of nuclear forces which is manifested in
the experimental fact that the volume of nucleus is roughly
proportional to the mass number A. However, the formula (1) is
not valid any more for nuclei in which numbers of protons ($Z$)
and neutrons ($N$) differ significantly.

It was found in Ref. \cite{ne} that the isospin dependent
formula for the nuclear charge radius constant
\begin{equation}
  r^{ch}_0 = 1.25 \left(1 - 0.2 \cdot {N-Z\over A}\right)~{\rm fm}\,,
%2
\end{equation}
describes much better than Eq. (1) the experimentally known
charge mean square radii of even-even nuclei with $A\geq 60$. The formula
(2) was obtained assuming the uniform charge distribution within
deformed nucleus.
In such an approximation the root mean square radius
(RMSR) of deformed
nucleus is given by the following formula
\begin{equation}
<r^2>^{1/2} = \sqrt{3\over 5} R_0\cdot g(\varepsilon,\varepsilon_4)\,,
%3
\end{equation}
where the function $g$ describes the dependence of the mean square radius
on deformation.
The equilibrium deformations of nuclei were taken from Ref.
\cite{ner}, where
the two-dimensional ($\varepsilon$, $\varepsilon_4$) space of
deformation parameters was used. The potential energy surfaces were calculated in
Ref. \cite{ner} by the Strutinsky prescription with the zero--point energy
correction terms according to the generator coordinate method \cite{nerl}.

Lateron it was found, after more broad calculations with the equilibrium
deformations taken from Ref. \cite{mol}, that the additional term
$\sim{1\over A}$ in the formula (2) should appear when
reproducing the mean square radii of all nuclei beginning from the lightest
ones up to the actinides \cite{nerlo}
\begin{equation}
 r^{ch}_0 = 1.240 \left(1  - 0.191\cdot {N-Z\over A}  + 1.646\cdot {1 \over A}
      \right)~{\rm fm}\,.
%3
\end{equation}

A further development of the formula for nuclear radius was made in Ref.
\cite{dob} after
extensive Hartree-Bogolubov calculations with various sets of Skyrme
effective interactions for proton and neutron density
distributions in a~few spherical nuclei. The authors of Ref. \cite{dob}
postulate  new terms proportional
to $1/A^2$ in the formula for $r^{ch}_0$.

The aim of our present work is to find simple formulae which
approximate the results obtained for
proton and neutron radii within the RMFT \cite{ser,rin}.
We investigate all even-even $\beta$ stable
nuclei as well as the isotopes and isotones corresponding to major magic $Z$
and $N$ numbers. When discussing the nuclides far from stability we change
the mass number $A$ up to the zero neutron or proton separation
energy.

In Section~2 we describe briefly the theoretical model and its
parameters.  In Section~3 we analyze the results of the calculations
and we write formulae for the radii of the proton and neutron
distributions, which approximate the theoretical results of
the RMFT.  A~simple, phenomenological formula for neutron radii is
proposed as the result of the above analysis.  The conclusions
and further investigations perspectives are described in
Section~4.

\section{Theoretical model}

The relativistic mean field theory \cite{ser} is a variational model
based on a standard Lagrangian density \cite{rin}
\begin{eqnarray}
%{rl}
{\cal L}
 &=& \bar\psi_i\left[\gamma^\mu(i\partial_\mu-g_\omega\omega_\mu-
 g_\rho\vec\rho_\mu
 \vec\tau-e\frac{1-\tau_3}{2}A_\mu)-M-g_\sigma\sigma\right]\psi_i \nonumber\\
 &+&{1\over 2}(\partial\sigma)^2- U(\sigma)
 -{1\over 4}\Omega_{\mu\nu}\Omega^{\mu\nu}
 +{1\over 2}m^2_\omega\omega^2 \\
 &-&{1\over 4}\vec R_{\mu\nu}\vec R^{\mu\nu}
 +{1\over 2}m_\rho^2{\vec\rho}^2
 -{1\over 4} F_{\mu\nu}F^{\mu\nu}\,\,,\nonumber
%4
\end{eqnarray}
consisting of nucleon $\psi$, mesons $\sigma, \omega, \vec\rho$
and electromagnetic $\vec A$ fields. The $\sigma$ mesons potential has been
taken in the nonlinear form:
\begin{equation}
U(\sigma)= {1\over 2}m^2_\sigma\sigma^2+{1\over 3}g_2\sigma^3+
           {1\over 4}g_3\sigma^4 \,\,.
\end{equation}
It was found in \cite{pom} that the NL-3 \cite{lal} parameters set of the
mean field Lagrangian (5)  reproduced well binding energies,
proton and neutron separation energies, electric quadrupole moments
and radii of all nuclei along the whole $\beta$--stability line. 
The NL-3 parameters are:

\begin{tabular}{ll}
-- nucleon mass & $M = 939$~MeV \\
-- meson masses & $m_\sigma =$ 508.194 MeV, $m_\omega = $ 782.501 MeV, \\
                &  $m_\rho =$ 763 MeV   \\
-- meson coupling constants & $g_\sigma=$ 10.217, $g_\omega$=
   12.868, $g_\rho $ = 4.474\\
-- $\sigma$ meson field constants & $g_2 = $ --10.431 1/fm, $g_3$= -28.885
\end{tabular}
\vspace{5mm}

The relativistic Hartree equations are solved by iterations: 
one starts with some estimate of the
meson and electromagnetic fields, then solving the Dirac equation one
finds the Dirac spinors. They give the densities and currents
as sources for the Klein-Gordon equations for the meson fields. After their
solution the new set of the meson and electromagnetic fields is found as the
starting point for the next iteration. When the selfconsistency is
achieved, Hartree-Bogolubov wave functions $\Psi^{p(n)}$ of protons and
neutrons are used for evaluation of the mean values of
operators in interest.
We assume here the product of the BCS--type functions
\begin{equation}
|\Psi\rangle = \prod_{\nu > 0} (u_\nu + v_\nu a^+_\nu a^+_{-\nu})
               |vac\rangle
%5
\end{equation}
for protons and neutrons as the ground state wave function of nucleus.

To get the strength of the pairing
interaction for nuclei from very different regions, we have taken simply
the experimental energy gaps ($\Delta$) and the lowest in energy $Z$ (or $N$)
single--particle levels
when solving the BCS equations. Such a procedure is justified in our
calculations because the quantities which we evaluate depend rather weakly
on the choice of the pairing force. All $\Delta$--s are extracted from the
experimental mass differences taken from the Wapstra--Audi tables \cite{wap}
but for the isotopes for which the experimental data do not exists we have
used the estimates: $\Delta_{p(n)}=12 / \sqrt{A}$~MeV.

The monopole moments of protons and neutrons distributions are
\begin{equation}
Q_0^{p(n)} = \langle\Psi|\sum_\nu r^2_\nu |\Psi\rangle^{p(n)}\,.
%6
\end{equation}
The mean square radii (MSR) are defined
\begin{equation}
 \langle r^2\rangle_p = {Q^p_0\over Z}\,,~~~~~~~~~
 \langle r^2\rangle_n = {Q^n_0\over N}\,.
%7
\end{equation}
and the root mean square radii  (RMSR) are
\begin{equation}
r_{p(n)} = \sqrt{\langle r^2\rangle_{p(n)}} = \sqrt{3 \over 5} R_{p(n)} \,.
%8
\end{equation}
In equations (9--10) we have neglected corrections originating
from the center of mass motion. For heavier nuclei which we
discuss these corrections are small.  The quadrupole moments of
proton and neutron distributions are given by
\begin{equation}
 Q^{p(n)}_2 = \langle\Psi|\sum_\nu 2r_\nu^2\,P_2(\cos\vartheta_\nu)
              |\Psi\rangle^{p(n)}\,,
%9
\end{equation}
where $P_2$ is the Lagrange polynomial of the order 2. The quadrupole
deformation parameter of proton or neutron distribution is approximately 
equal to
\begin{equation}
 \beta^{p(n)}_2 \approx \sqrt{{4\pi\over 5}} {Q^{p(n)}_2 \over Q^{p(n)}_0}\,.
%10
\end{equation}
We have evaluated also the binding energies of nuclei ($B_{\rm RMF}$), 
the reduced electric quad\-ru\-po\-le transition probabilities ($B(E2)$), 
the proton and neutron separation energies ($S_{p(n)}$).
The results are compared with the experimental data taken from Refs.
[1, 2, 14].

\section{Results}

Various functions of radii and density moments were investigated in order to
extract the isospin dependence of proton and neutron density distributions.
The calculations were performed for the stable nuclei along the $\beta$
stability line and for all potentially existing isotopes with
$Z$=50, 56, 82, 94 and isotones with $N$=50, 82, 126.
Schematic view in the ($N,Z$) plane of the nuclides discussed
in the paper is presented in Fig. 1. For the nuclei out of
$\beta$--stability line the calculation was made until the proton
or neutron drip line was reached.

In Fig. 2 the set of results concerning  the $\beta$-stable nuclei
with $40 \leq A \leq 256$ is presented as a function of mass number $A$.
For each $A$ value only one isotope with the  smallest mass is
chosen.

The upper left figure shows the difference  between the binding energy 
$B_{\rm RMF}$ of a~nucleus calculated by the RMFT with the NL-3 parameters 
set and its experimental value $B_{\rm exp}$ taken from \cite{wap}.
Here the maximal error of the RMFT estimates is 5~MeV only (for $A=150$).

The upper right figure presents the reduced electric
quadrupole transition probabilities $B(E2)$ obtained by the RMFT (solid line)
compared to the experimental data \cite{ott} (circles). The agreement 
is rather good in spite of too large theoretical predictions of the
nuclei about $A \sim 170$ and 250.

The separation energies of neutron $S_n$ (left hand side) and proton
$S_p$ (right hand side) evaluated within the RMFT agree with the
experimental data \cite{wap} (circles) very well.

The neutron $r_n$ and proton $r_p$ root mean square radii calculated
by the RMFT presented in the lowest left hand side figure
are also   close to the experimental data (crosses for
protons \cite{ott} , circles for neutrons \cite{batt}).
The both radii are slightly different from each other, what is connected
with the different in size and in deformation proton and neutron
density distributions.

The difference of the proton and neutron quadrupole deformations
obtained in the RMFT calculation is shown in the lowest right part of
Figure 2. One can see that for same nuclei it exceeds 0.03.

The next seven figure sets (Figs. 3-9) show  similar results as in
Fig. 2 but for the isotope and isotone chains presented in Fig. 1. The
experimental data (crosses for protons, circles for neutrons) are drawn, if
they exist. The results
should illustrate the goodness of the RMFT also for the nuclei
outside $\beta$ stability line. Using these results we can extract the 
isospin dependence of proton and neutron radius constants.

In Fig. 3 the same quantities as in Fig. 2 for Sn isotopes with
$N = 50-86$ are drawn. All  Sn  nuclei are almost spherical so the
quadrupole deformation of proton and neutron distribution is 0
(lowest right hand side picture). Also the reduced electric quadrupole 
transition probabilities $B(E2)$ proportional to the square of quadrupole 
moment are zero in the theory and the experiment as well (upper right hand 
side picture).
The results both for proton and neutron radii are well confirmed
by the experimental data what can be seen in the lowest left
figure. The RMFT radii of neutron and proton distributions
differ from each other up to 0.5~fm for the heaviest isotopes.
The experimental separation energies are very well reproduced by
the RMFT calculation (middle pair of figures).
The binding energies obtained by the RMFT differ only up to 2.5~MeV from
the experimental ones (upper left hand side picture).

The Ba isotopes presented in Fig. 4 are slightly deformed.
One can see in the lower right figure that the
neutron and proton deformations are slightly different.
The RMFT reproduces the experimental kink in neutron separation energy $S_n$
around neutron magic number $(N=82)$

The Pb isotopes are presented in Fig. 5. The  equilibrium deformations
are largest for he isotopes around $A=195$. The proton and neutron radii 
differ from each other much especially for the lighter isotopes.
Unfortunately the RMFT estimate of the
neutron root mean square radius for $^{208}$Pb is pretty far from the
experimental value \cite{batt}.

The Pu isotopes, shown in Fig. 6 behave quite regularly in
spite of the relative large deformations and $B(E2)$.
The quadrupole deformations of
proton and neutron distributions differ by about 0.025 for all
isotopes. This difference is about 10 \% of the average  deformation of
these nuclei.

In spite of the constant $Z$ values the proton radius in Sn, Ba and Pb 
isotopes grows with the neutron number. It is not the case
for the isotones shown in the next 3 figures: 7--9, where the
neutron radii stay almost constant when number of protons $Z$
grows.

In Fig. 7 the $N=50$ spherical isotones  (Ni-Sn) are presented.
The errors in the binding energy are rather small. Due to the
spherical shape of these nuclei no differences in proton and
neutron deformations are observed. Contrary to the neutron
radius the proton radius grows here rapidly with $A$.

Also the $N=82$ isotones (Cd-Hf) shown in Fig. 8 have
an almost constant neutron density distribution radius.
This conclusion is also confirmed by the results presented in Fig. 9, for
$N=126$ isotones of Hg-U.

In order to extract the average isospin dependence of the radius
constant from the root mean square radii evaluated within the RMFT we have
divided the RMSR values by the factor $\sqrt{{3/ 5}}A^{1/3}$ (see Eq. 10).
As one can see in Figs. 10a and 10b the radius constants of protons (crosses)
and neutrons (circles) are gathered around 1.25~fm and 1.20~fm respectively 
as functions of the relative neutron excess $I=(N-Z)/ A$.
In order to analyze better the results one has to remove the influence of 
deformation on nuclear radii. We have renormalized them to the sphere using
the volume conservation rule. It was  made approximately by dividing the 
nuclear radii constants by factor $g$:
\begin{equation}
  g \left( \beta _2 \right) =  1 + \frac{5}{4 \pi }\beta _2^2\,.
%13
\end{equation}

This renormalized quantities $R_0 = r/(\sqrt{3/5}\cdot g( \beta _2))$
are shown on Fig. 10a,~b for Ba and Pb isotopes respectively.
Now one can see that the renormalized proton radii (diamonds)
decrease almost linear with I while the neutron ones (dots)
increase.

The above investigation has convinced us that the microscopic radius
constant $r^{p,n}_0$ must depend also on the isospin $I$. The additional
dependence on $1/A$ was noticed when comparing the results for different
isotope chains.
In Figs. 11, 12 the results are shifted by the terms
proportional to $1/A$ in order to see better the average $I$~dependence  
of the radii.

In Fig. 11 we present the proton and in Fig. 12 the neutron
renormalized radii constant for all the isotope and isotone
chains discussed in this paper (a) and for the $\beta$-stable nuclei
(b). The results for protons are shifted by the term $0.8 /A$ fm while
the neutron estimates by  $-3.3 /A$ fm.

The dashed lines present average behavior of calculated radii constants. 
It was found by minimization of the mean square deviations. The following
formulae for proton and neutron radii:
\begin{eqnarray}
&& R^p_0 = 1.237 \left(1 - 0.157 \cdot {N-Z\over A} - 0.646 \cdot \frac{1}{A}
   \right)A^{1/3}~{\rm fm}\nonumber\\
&&\hspace{-3.5cm}{\rm and}\\
&&  R^n_0 = 1.176 \left(1 + 0.250 \cdot {N-Z\over A}+ 2.806 \cdot \frac{1}{A}
\right) A^{1/3}~{\rm fm}\nonumber
%14
\end{eqnarray}
approximate the results obtained within the RMFT with the NL-3
set of the parameters.  One can believe that the parameters of these
formulae, found for the representative nuclei all over the
periodic system can describe well the average trend of the
results obtained within the RMFT.

The following formula approximates the RMFT results for the nuclear charge 
radii:
\begin{equation}
 R^{ch}_0 = 1.241 \left(1 - 0.154 \cdot {N-Z\over A} + 0.580\cdot \frac{1}{A}
\right)
A^{1/3}~{\rm fm}\,.
%15
\end{equation}
This equation, obtained only by the analysis of the RMFT calculation
results, has not much different parameters to those from equation
(3) fitted to all available experimental data. The coefficient
of $(N-Z)/ A$ term is only slightly smaller than in the
phenomenological formula (3). But the parameter at the term
$1/A$ in (15) is almost 3 times smaller than in Eq. (3). It is mostly
due to the fact that we have analyzed the nuclei with $A\geq 40$ while 
the Eq. (3) was obtained by fitting the experimental data for all nuclei with
$A\geq12$.

From Figs. 11 and 12 we can learn that the average formulae (14) work 
properly only for nuclei with $A\geq 60$. Also some
nuclei with $I\sim 0.2$ don't fit  to  the average formulae. It is caused by
non quadrupole deformations (octupole?) of these nuclei which were not 
included in the present analysis.

As far as the ratio of proton to neutron radii (or RMSR) is
concerned there is almost no influence of deformation on the
results. It is caused by the similar or identical shapes of proton
and neutron density distributions. We have found a~short
formulae for this ratio similar to those for radii constants (14,15). 
The dependence of proton to neutron radii ratio on $I$ is shown in Fig. 13.  
We have shifted the ratio by the term $-3.3/A$ in order to remove
the influence of the $1/A$ dependence. Dashed lines in Fig. 13
represent the formula:
\begin{equation}
 \frac{r_p}{r_n} = 1.048 \left(1 - 0.364 \cdot {N-Z \over A} - 3.148\cdot
  {1 \over A} \right)
%16
\end{equation}
\noindent
fitted by the least square fit to the $r_p/r_n$ RMFT ratios 
for all nuclides shown Fig. 1. In Fig. 13a the radii
ratios of these chains of isotopes and isotones are presented while in 
Fig. 13b the corresponding results for $\beta$-stable nuclei with $A\geq 40$ 
are plotted. The results for nuclei with $A\geq 60$ from
are reproduced well by formula (16).

Eq. (16) can be used to estimate the neutron RMSR from the charge one.
It could be useful because it exists over 250 measured charge RMSR and only  
few experimental ones for the neutron RMSR. One can rearrange Eq. (16) 
using the relation between the charge and proton mean square radii
\begin{equation}
 r_{ch}^2 = r_p^2 + 0.64~{\rm fm^2} \,.
%17
\end{equation}
The term 0.64 fm$^2$ originates from the final size of the proton.
Finally one gets:
\begin{equation}
r_n =  {( r_{ch}^2 - 0.64)^{1/2} \over
1.048 \left( 1 - 0.364 \cdot (N-Z)/ A - 3.148 / A\right)}~{\rm fm}\,.
%18
\end{equation}

In Fig. 14 all known experimental RMSR (errorbars)
\cite{batt} for neutrons are compared  with the RMFT predictions
(crosses) and with the results estimated from experimental charge
RMSR \cite{batt} using Eq. (18) (circles). All results agree very well with 
each 
other, so we hope that the formula (18) can be used by the experimentalists 
to foresee the neutron distribution radii for other nuclei as well.

\section{Conclusions}

The following conclusions can be drawn from our calculation
\begin{itemize}
\item[1.] The renormalized to sphere RMFT proton and neutron distributions
          radii $R^{p(n)}_0$ depend almost linear on neutron excess 
          $I=(N-Z)/ A$. $R^p_0$ decrease with $I$, while 
          $R^n_0$ increase what was suggested in \cite{nerlo} and obtained
          in \cite{dob} by the analysis of the results obtained within
          the HFB calculation with the Skyrme forces.

\item[2.] The parameters of the formulae for the charge radii are similar 
          as in the phenomenological formula in \cite{nerlo}. However the
          dependence of the RMFT radii on $I$ is about 20\% weaker than
          in the phenomenological formula which describes the global 
          experimental trend. It means that the parameters NL-3 of the 
          RMFT should be slightly changed. 

\item[3.] The term  $\sim A^{-1}$ in  the formula for $R^{p(n)}_0$ is 
          needed in order to reproduce the average MSR values obtained 
          within the RMFT. Its role is especially important for lighter
          nuclei. 

\item[4.] The ratio of the proton radius to the neutron radius is a smooth
          function of $I$ and $A^{-1}$ and it could be very well described
          by the simple formula (16). Using this global dependence we have 
          written the phenomenological formula (18) which allows to foresee
          the magnitude of the neutron radius when the experimental charge
          radius is known. The prediction power of the formula (18) is not
          worse than that of very advanced microscopical calculations
          based on the RMFT or the HFB-Skyrme model \cite{dob}.

\end{itemize}

The obtained formulae for the $R^n_0$ and $R^p_0$ radii will be used to
develop the liquid drop like model, which will depend on the different 
proton and neutron density distributions, i.e. different radii and 
deformations. Such investigations are in progress now \cite{die}.

\vspace{1cm}

\noindent
{\large\bf Acknowledgment:}
\vspace{0.5cm}

The authors gratefully acknowledge the helpful discussions with Peter Ring
and offering by him the numerical code for solving the relativistic
Hartree model.

\newpage

\centerline{Figure captions}

\bigskip
\begin{enumerate}
%1
\item Schematic view in the ($N,Z$) plane of the nuclides discussed
      in the paper. For the nuclei out of the $\beta$-stability line
      the calculation was made until the proton or neutron drip line
      was reached.
%2
\item The results obtained within the RMFT+BCS model with the NL-3
      parameters \cite{lal} for the $\beta$-stable nuclei compared with
      the experimental data as a functions of mass number A. In the
      diagrams are presented: the difference between the calculated
      $B_{\rm RMF}$ and experimental $B_{\rm exp}$ binding energy
      (upper-left figure), reduced electric
      quadrupole transition probabilities $B(E2)$ (upper-right figure),
      neutron and proton separation energies $S_n, S_p$ (middle figures),
      neutron and proton root mean square radii $r_n, r_p$ (lower-left
      figure), differences between proton and neutron quadrupole deformation
      parameters $\beta_2^p - \beta_2^n$ (lower-right figure).
%3
\item Same as in Fig. 2 but for the Sn isotopes ($Z=50$).

%4
\item Same as in Fig. 2 but for the Ba isotopes ($Z=56$).

%5
\item Same as in Fig. 2 but for the Pb isotopes ($Z=82$).

%6
\item Same as in Fig. 2 but for the Pu  isotopes ($N=94$).

%7
\item Same as in Fig. 2 but for the isotones with $N=50$.

%8
\item Same as in Fig. 2 but for the isotones with $N=82$.

%9
\item Same as in Fig. 2 but for the isotones with $N=126$.

%10
\item The proton (crosses) and neutron (open circles) nuclear radii divided
      by $A^{1/3}$ as function of the relative neutron excess $I$ for Ba
      (a) and Pb (b) isotopes. Nuclear radii renormalized to sphere are 
      plotted with diamonds and dots respectively.

%11
\item Renormalized to sphere proton radii constants for discussed chains 
      of isotopes and isotones (a) and for all $\beta$ stable nuclei
      with $A\geq 40$ (b) as function of the relative neutron excess $I$
      shifted down by the $-3.3/A$ fm term.  Dashed line represents
      average behavior of calculated values for the chains (14). 

%12
\item Same as in Fig. 11 but for neutron radii (shifted up by the $0.8/A$ fm
      term).

%13
\item The ratios of the proton to neutron root mean square radii for 
      discussed chains of isotopes and isotones (a) and for the $\beta$ stable
      nuclei with $A\geq 40$ (b) as functions of relative neutron excess
      $I$. The ratios are shifted up by $3.3/A$ fm term. Dashed lines 
      represent the average behavior of calculated values (16).

%14
\item Comparison of the experimental neutron root mean square radii 
      \cite{batt} (errorbars) with the RMFT results (crosses) and the values
      obtained from experimental charge radii \cite{batt} through the new 
      formula (18) (circles).

\end{enumerate}

%\end{document}

\newpage
\begin{figure}
\epsfxsize=160mm \epsfbox{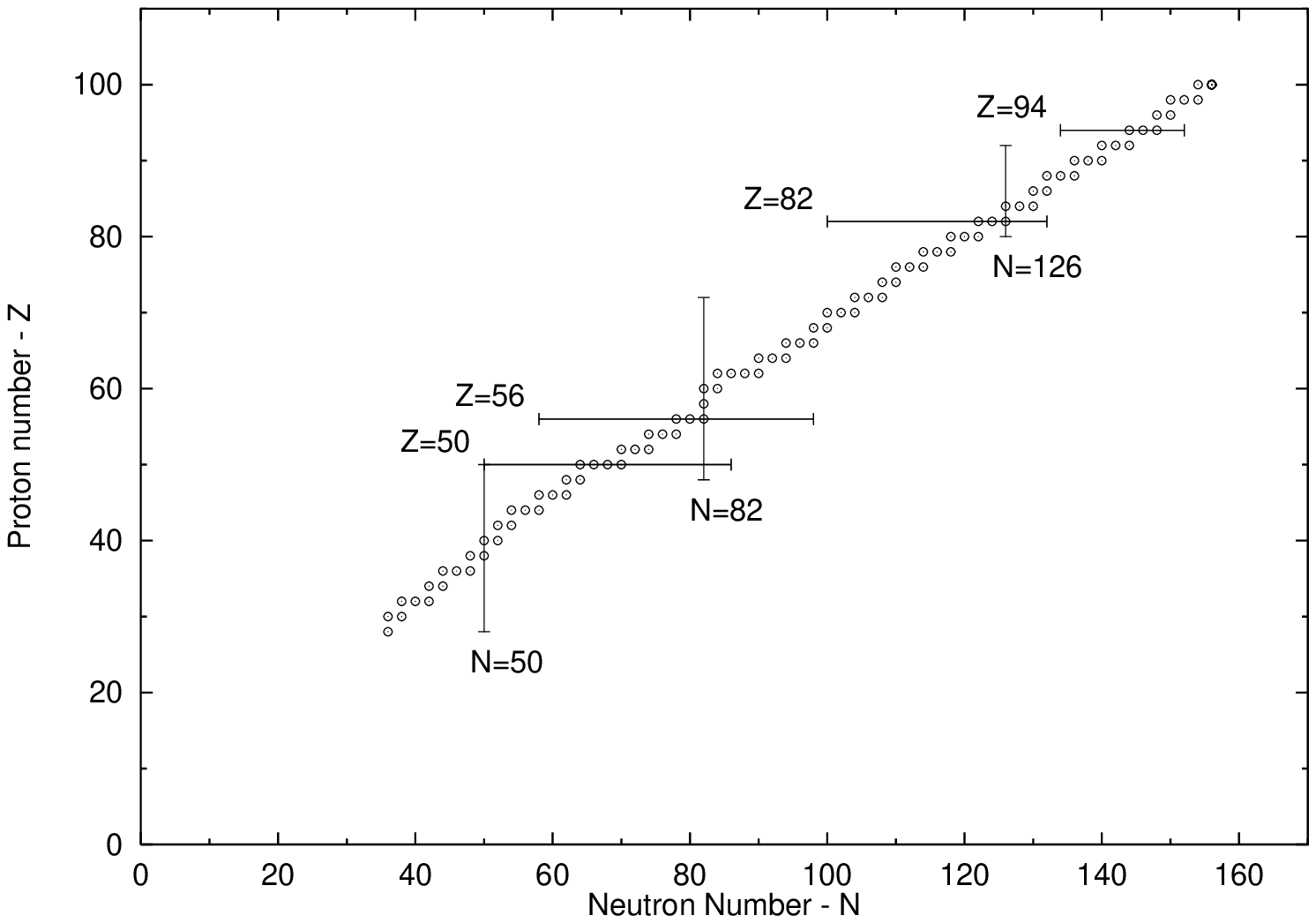}
\caption{ }
\end{figure}

\pagebreak[5]
\begin{figure} 
\epsfxsize=160mm \epsfbox{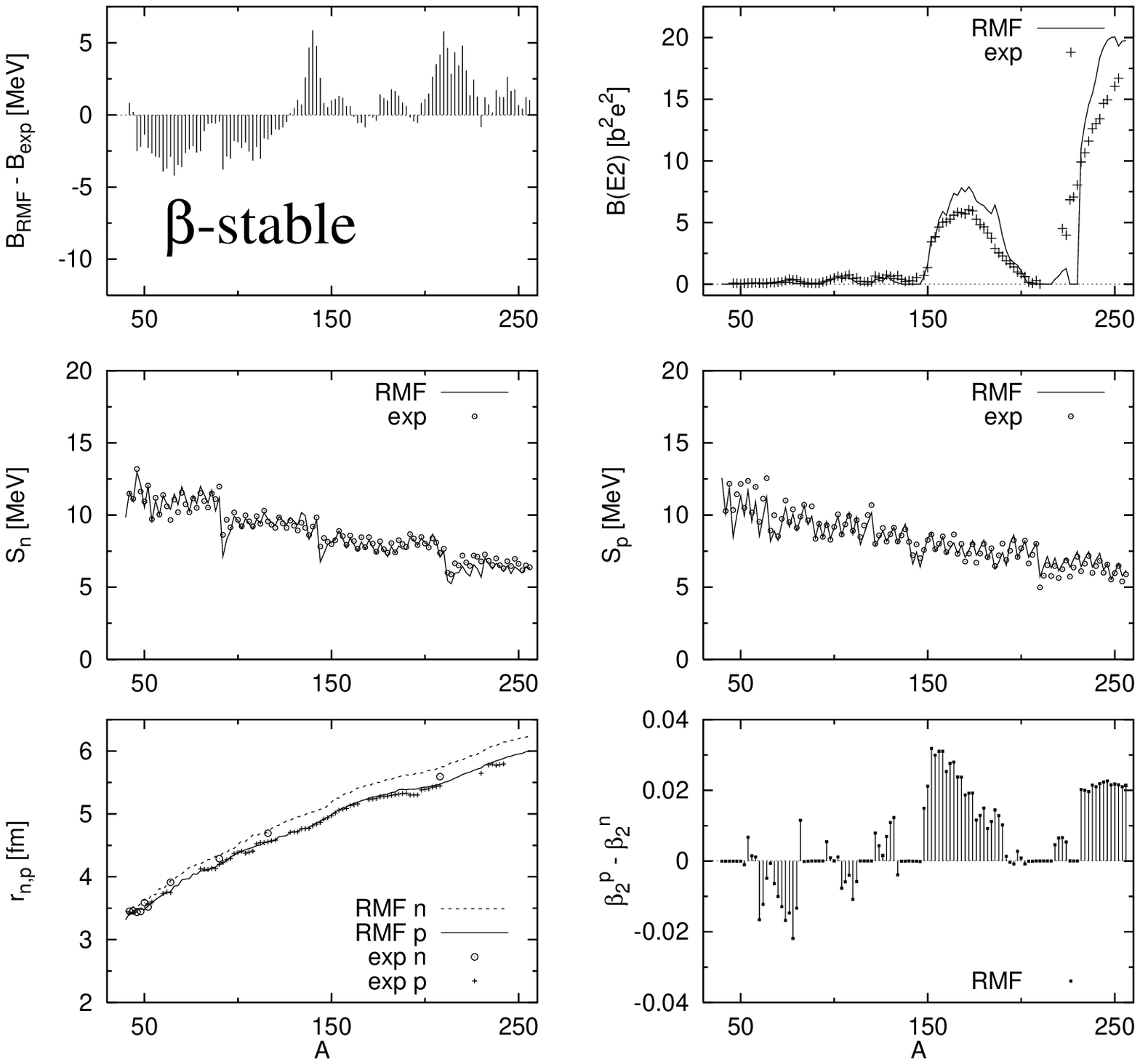}
\caption{ }
\end{figure}

\pagebreak[5]
\begin{figure}
\epsfxsize=160mm \epsfbox{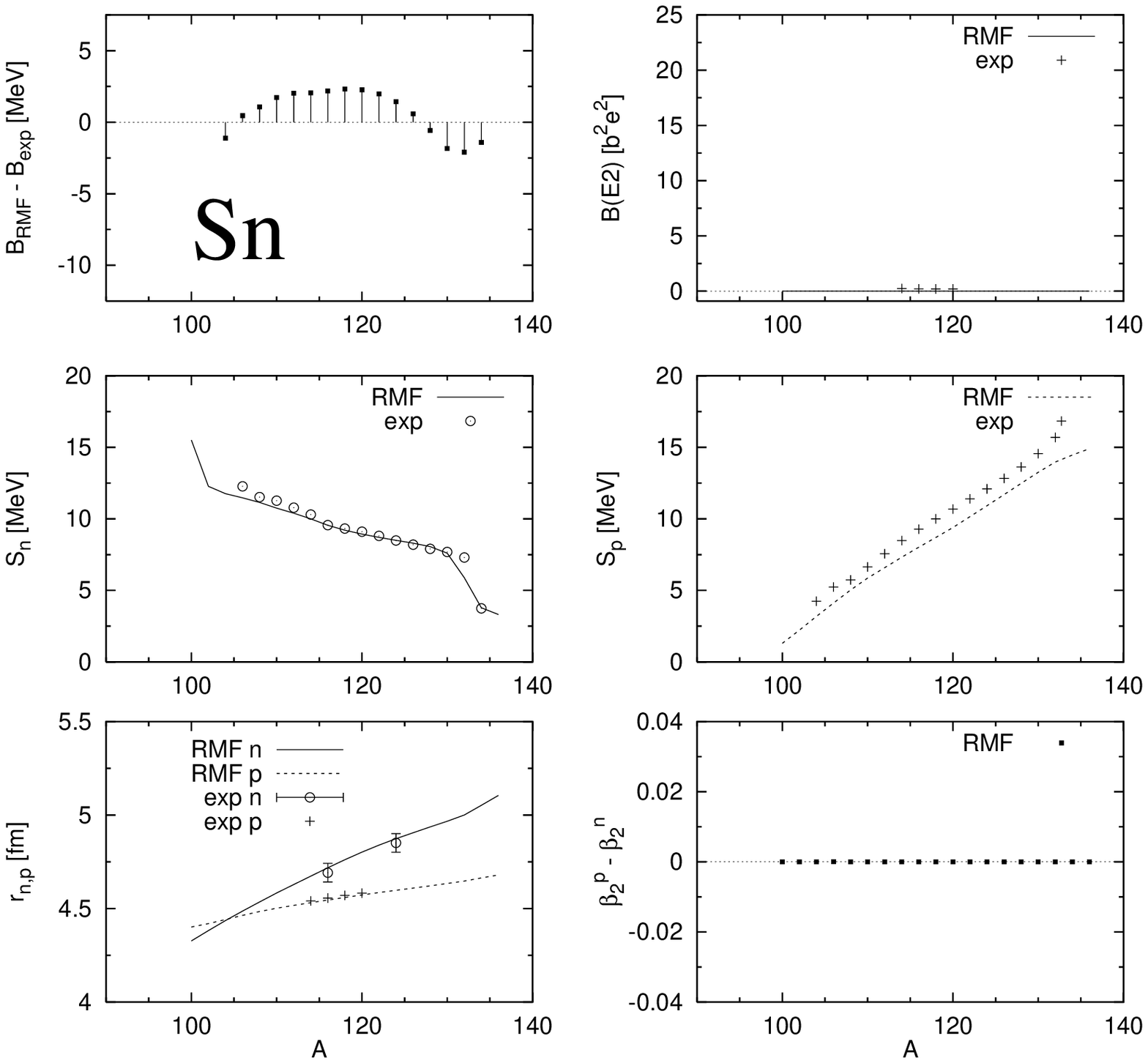}
\caption{ }
\end{figure}

\pagebreak[5]
\begin{figure}
\epsfxsize=160mm \epsfbox{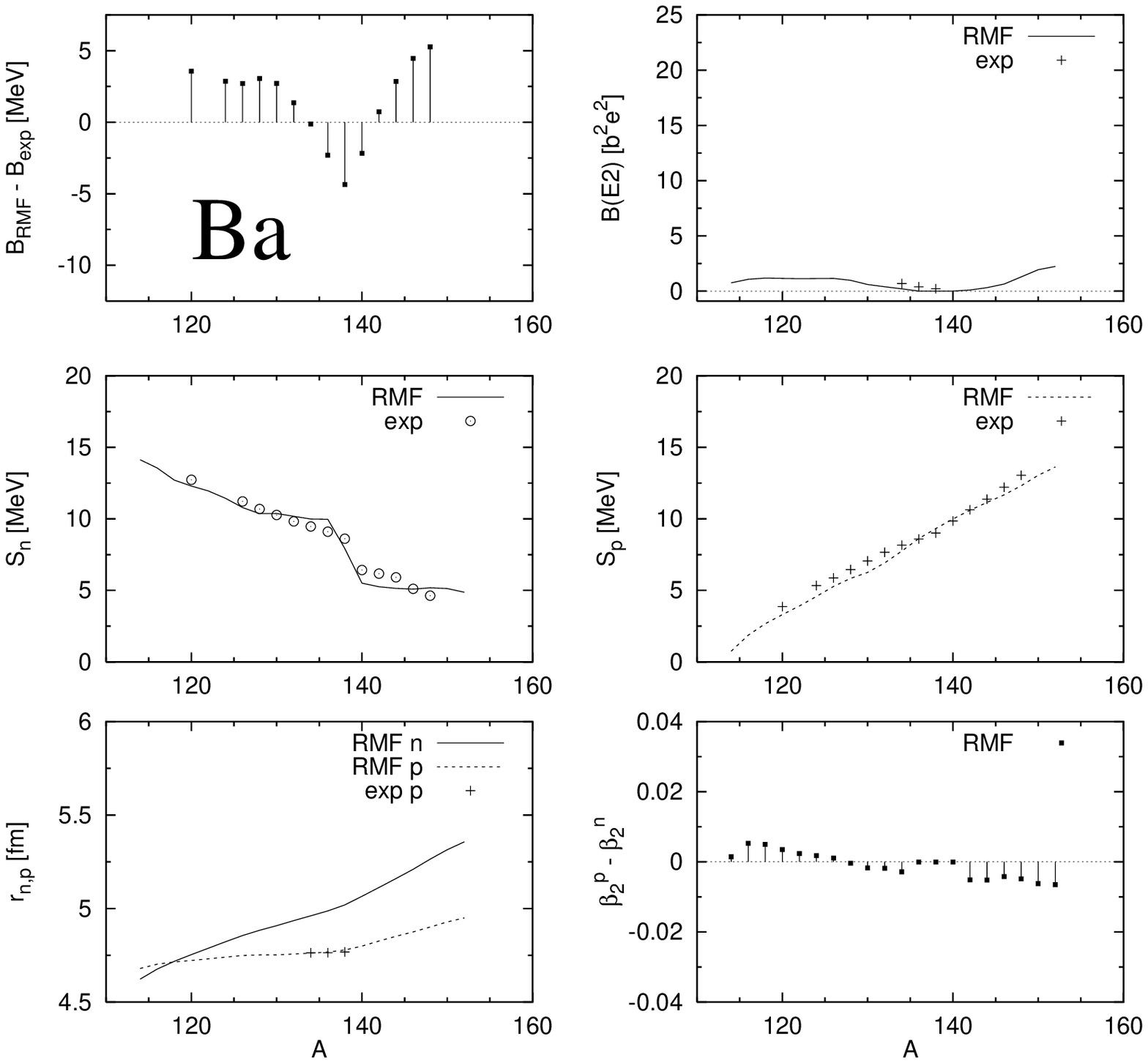}
\caption{ }
\end{figure}

\pagebreak[5]
\begin{figure}
\epsfxsize=160mm \epsfbox{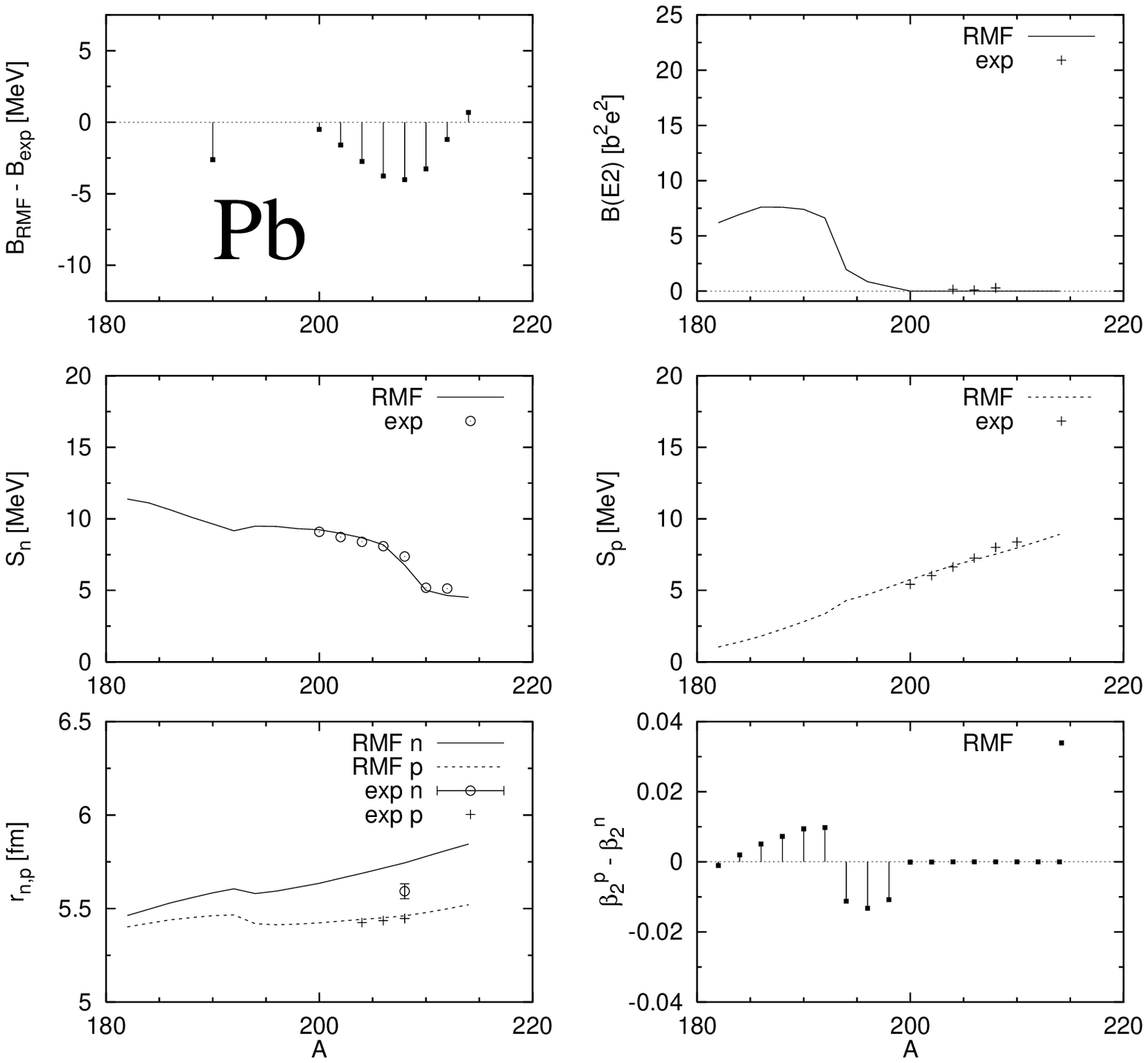}
\caption{ }
\end{figure}

\pagebreak[5]
\begin{figure}
\epsfxsize=160mm \epsfbox{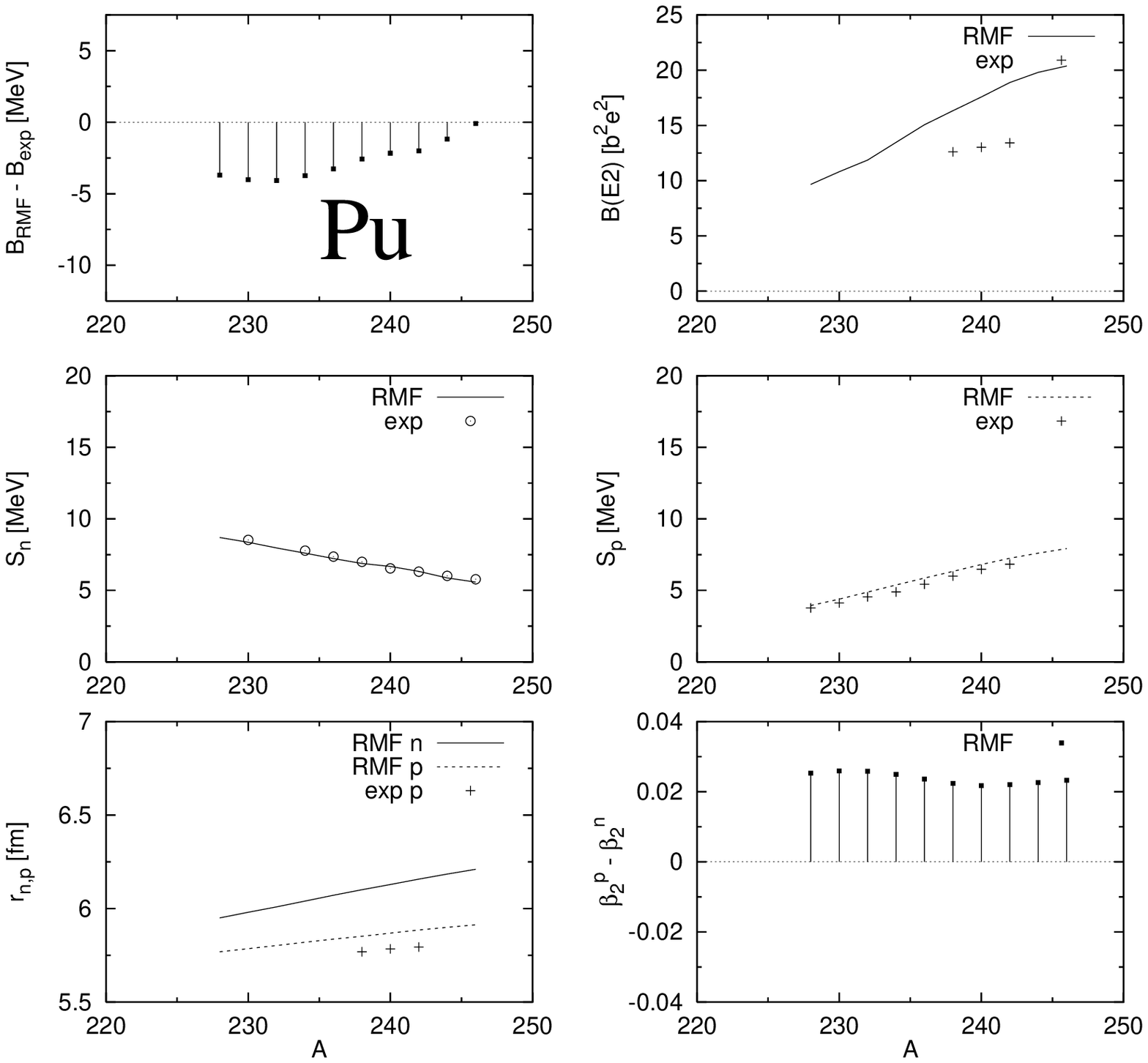}
\caption{ }
\end{figure}

\pagebreak[5]
\begin{figure} 
\epsfxsize=160mm \epsfbox{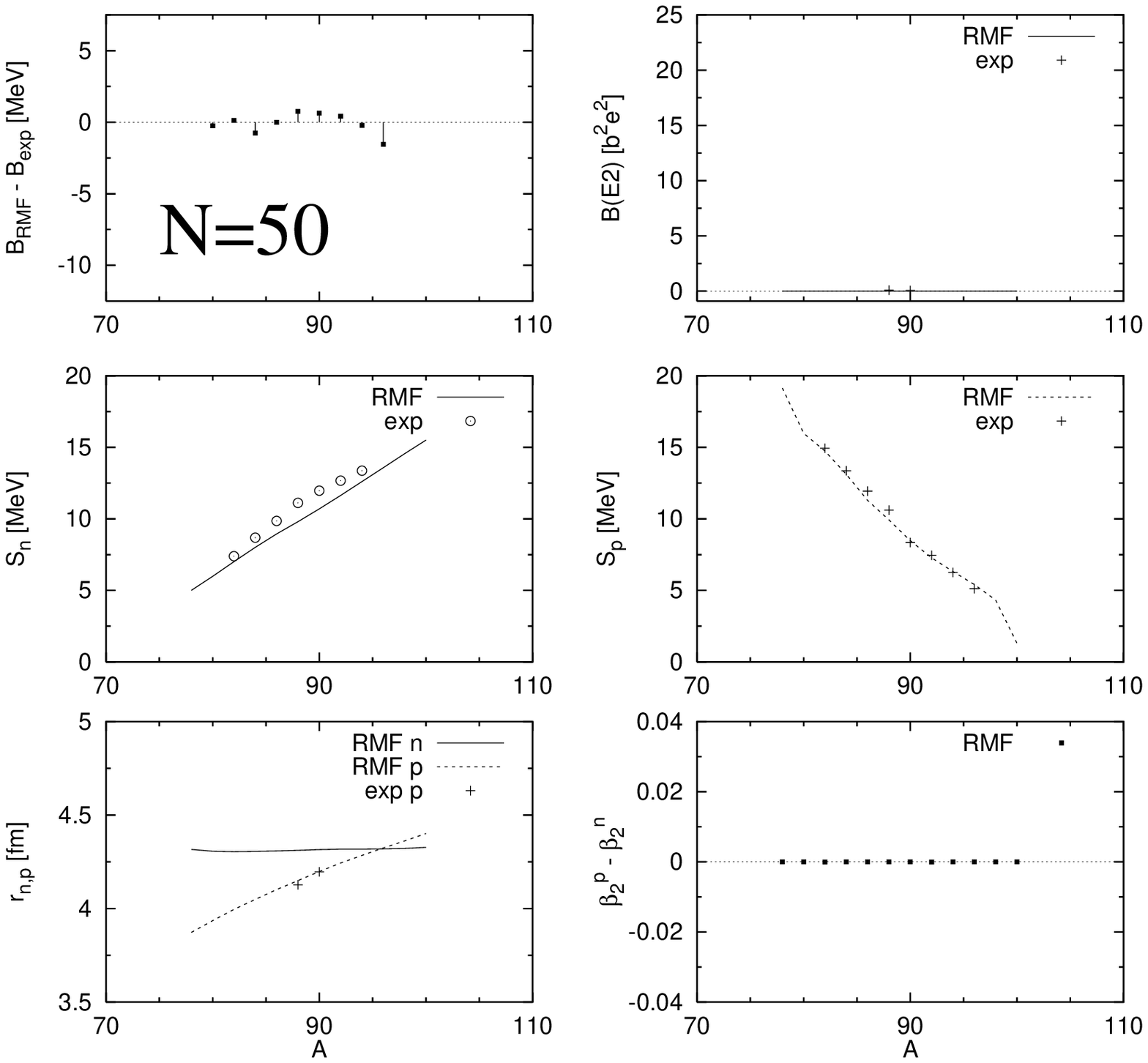}
\caption{ }
\end{figure}

\pagebreak[5]
\begin{figure}
\epsfxsize=160mm \epsfbox{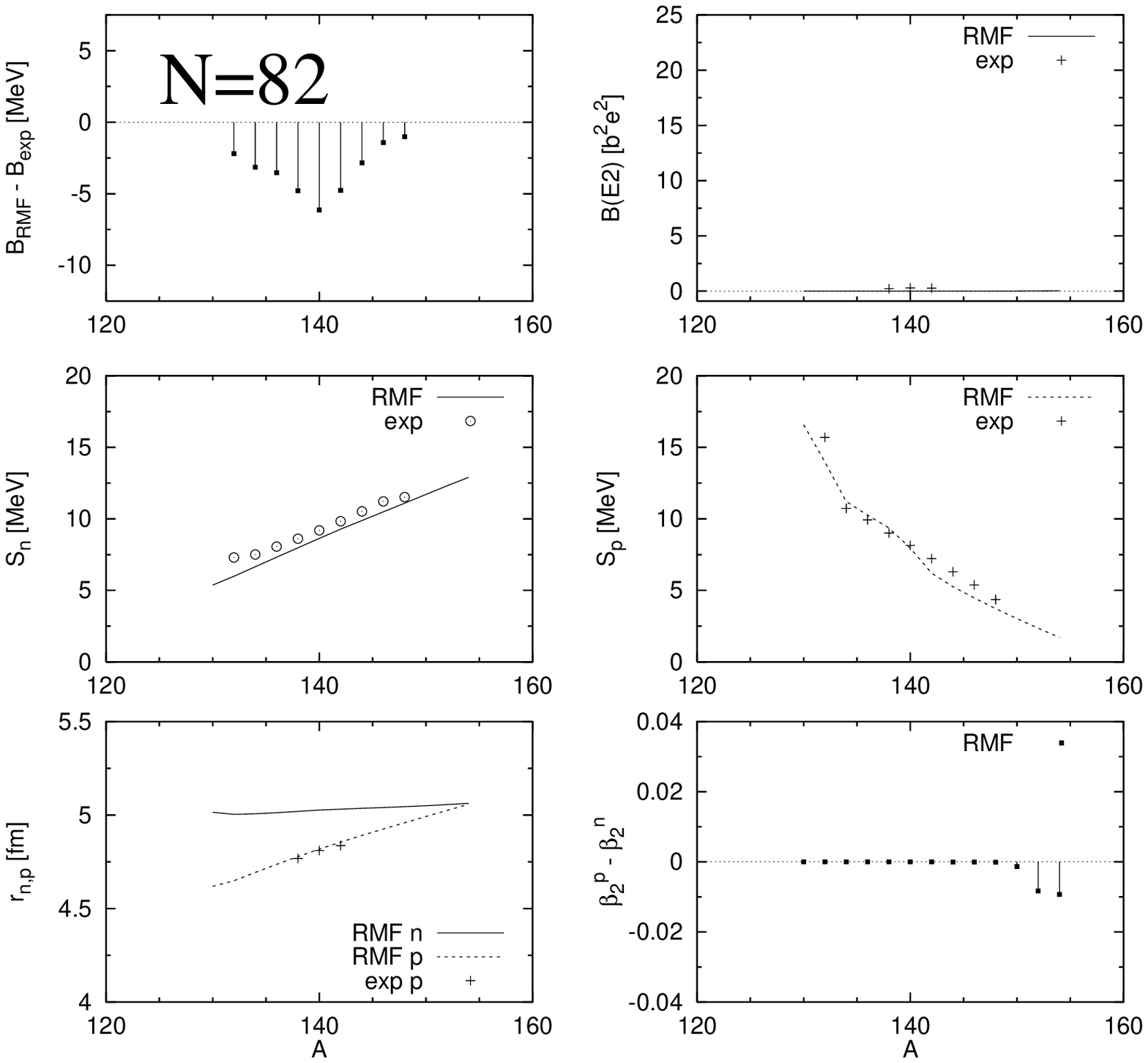}
\caption{ }
\end{figure}

\pagebreak[5]
\begin{figure}
\epsfxsize=160mm \epsfbox{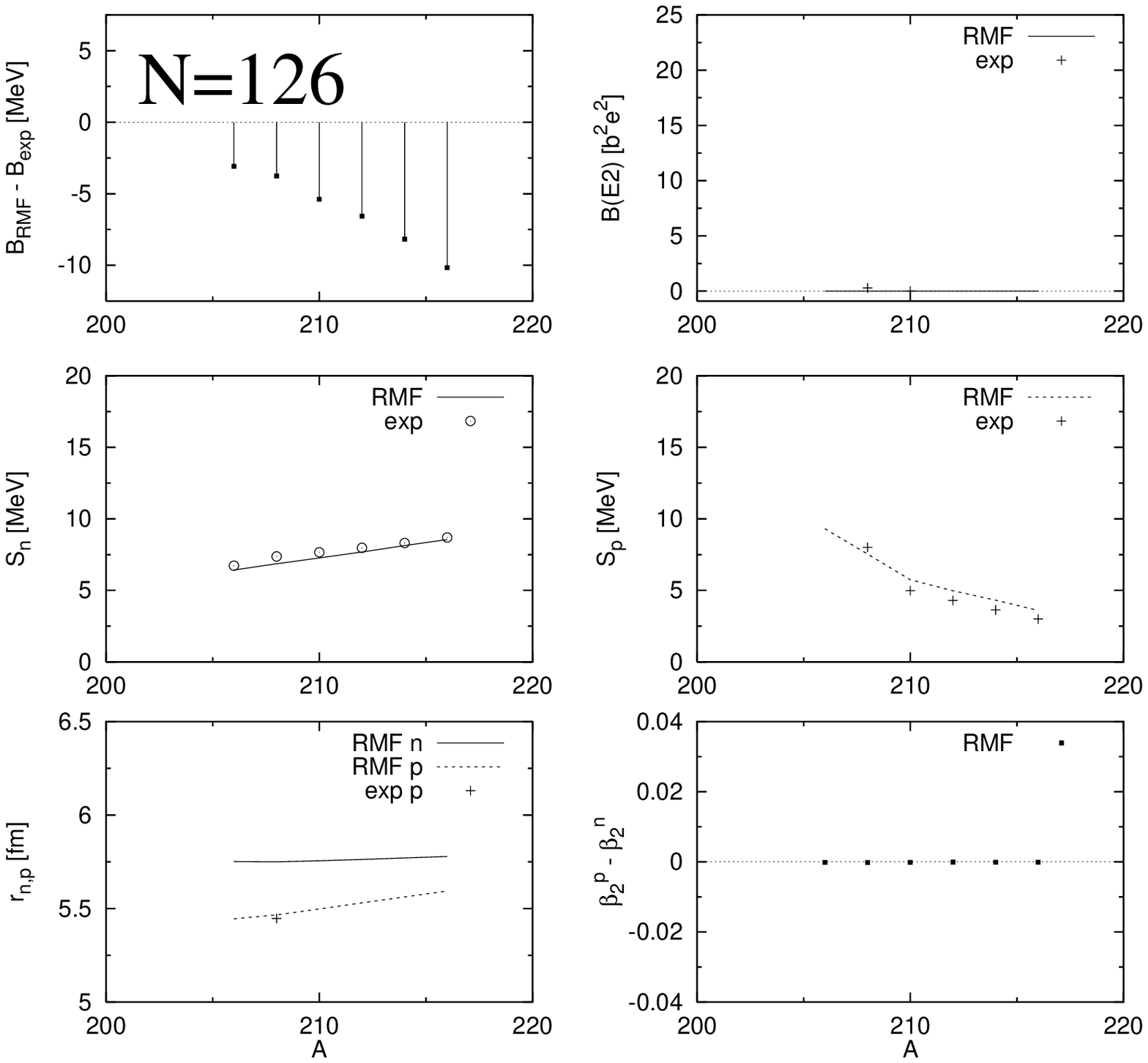}
\caption{ }
\end{figure}

\pagebreak[5]
\begin{figure}
\epsfxsize=160mm \epsfbox{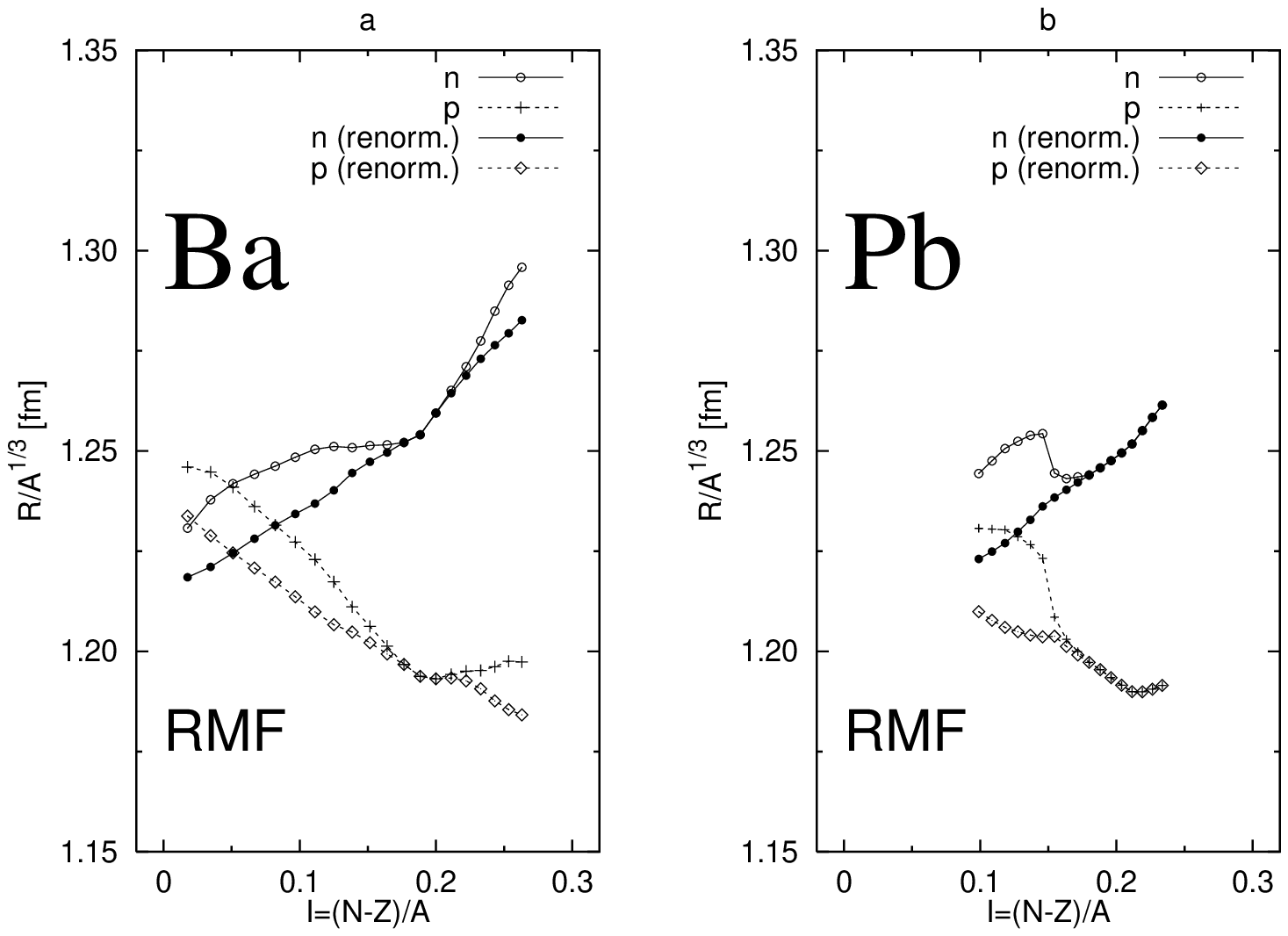}
\caption{ }
\end{figure}

\pagebreak[5]
\begin{figure} 
\epsfxsize=160mm \epsfbox{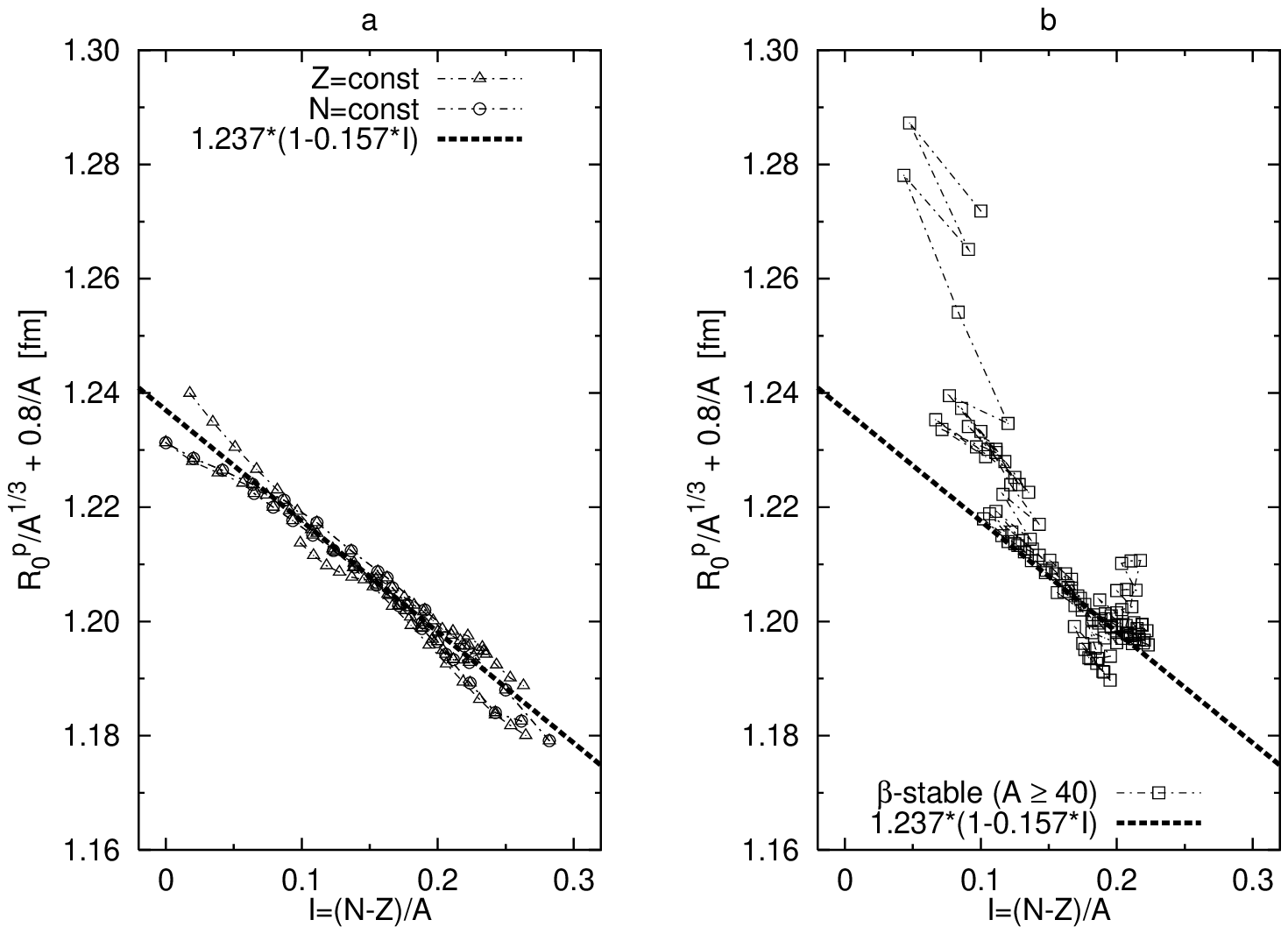}
\caption{ }
\end{figure}

\pagebreak[5]
\begin{figure} 
\epsfxsize=160mm \epsfbox{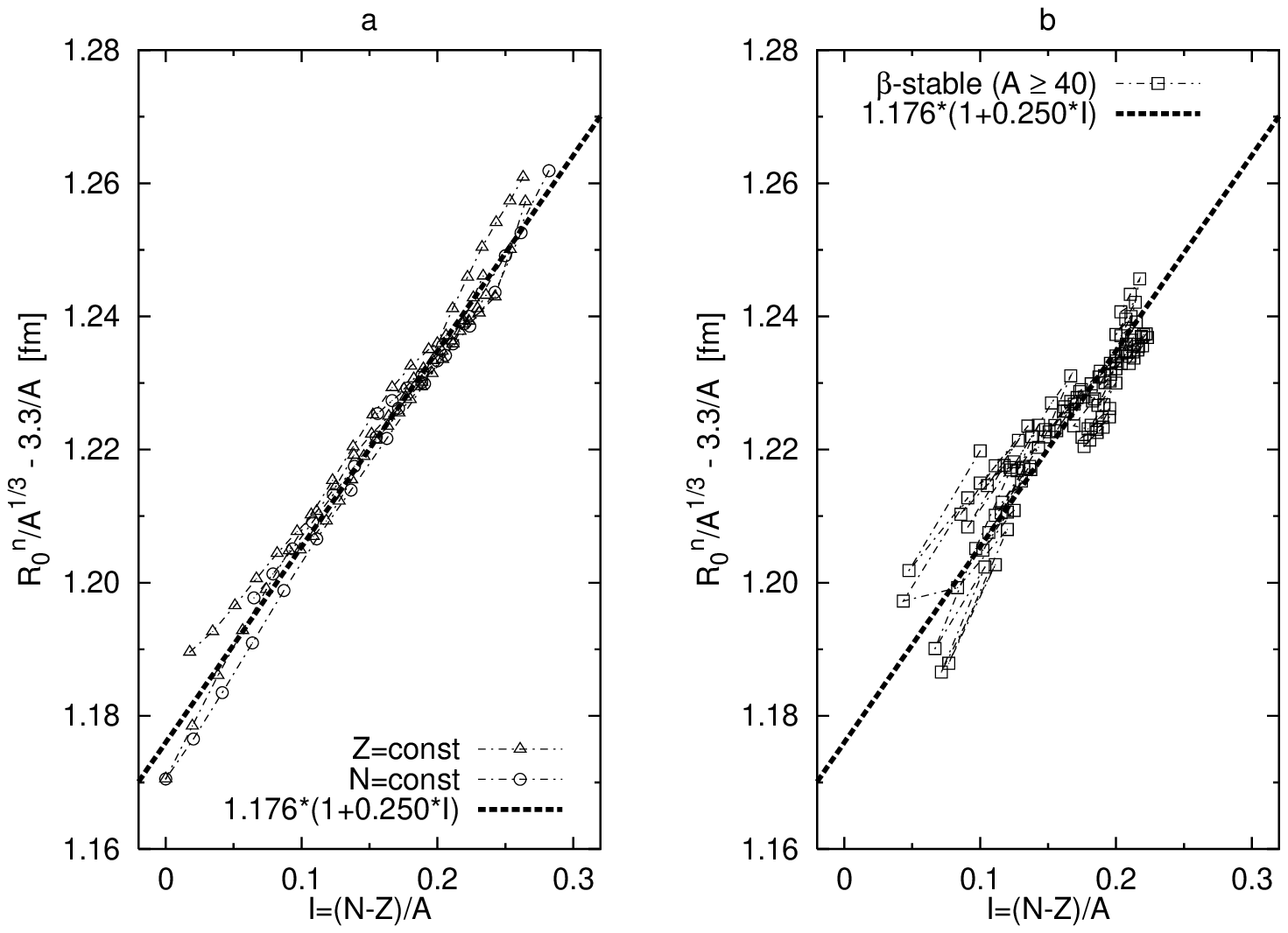}
\caption{ }
\end{figure}

\pagebreak[5]
\begin{figure}
\epsfxsize=160mm \epsfbox{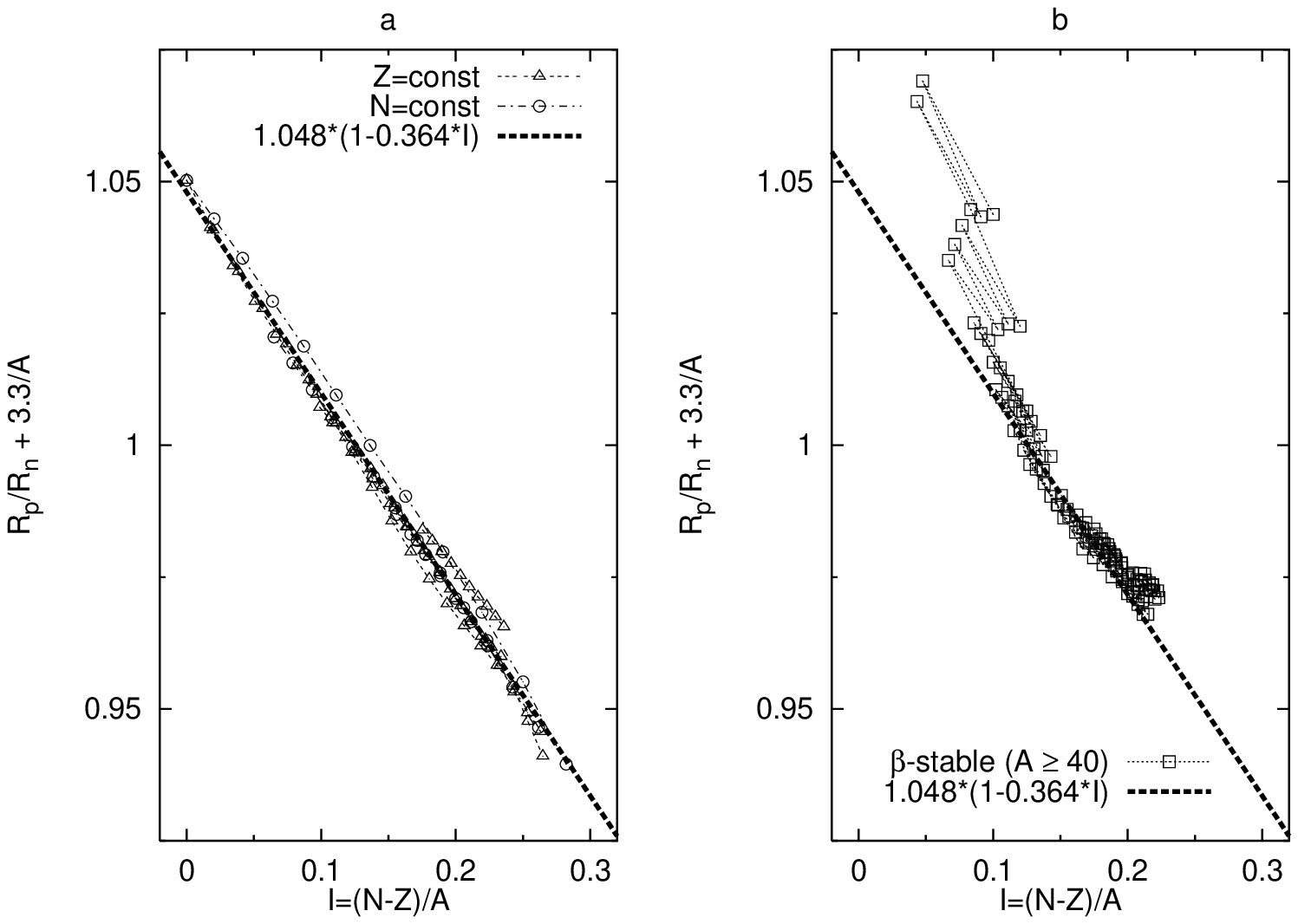}
\caption{ }
\end{figure}

\pagebreak[5]
\begin{figure}
\epsfxsize=160mm \epsfbox{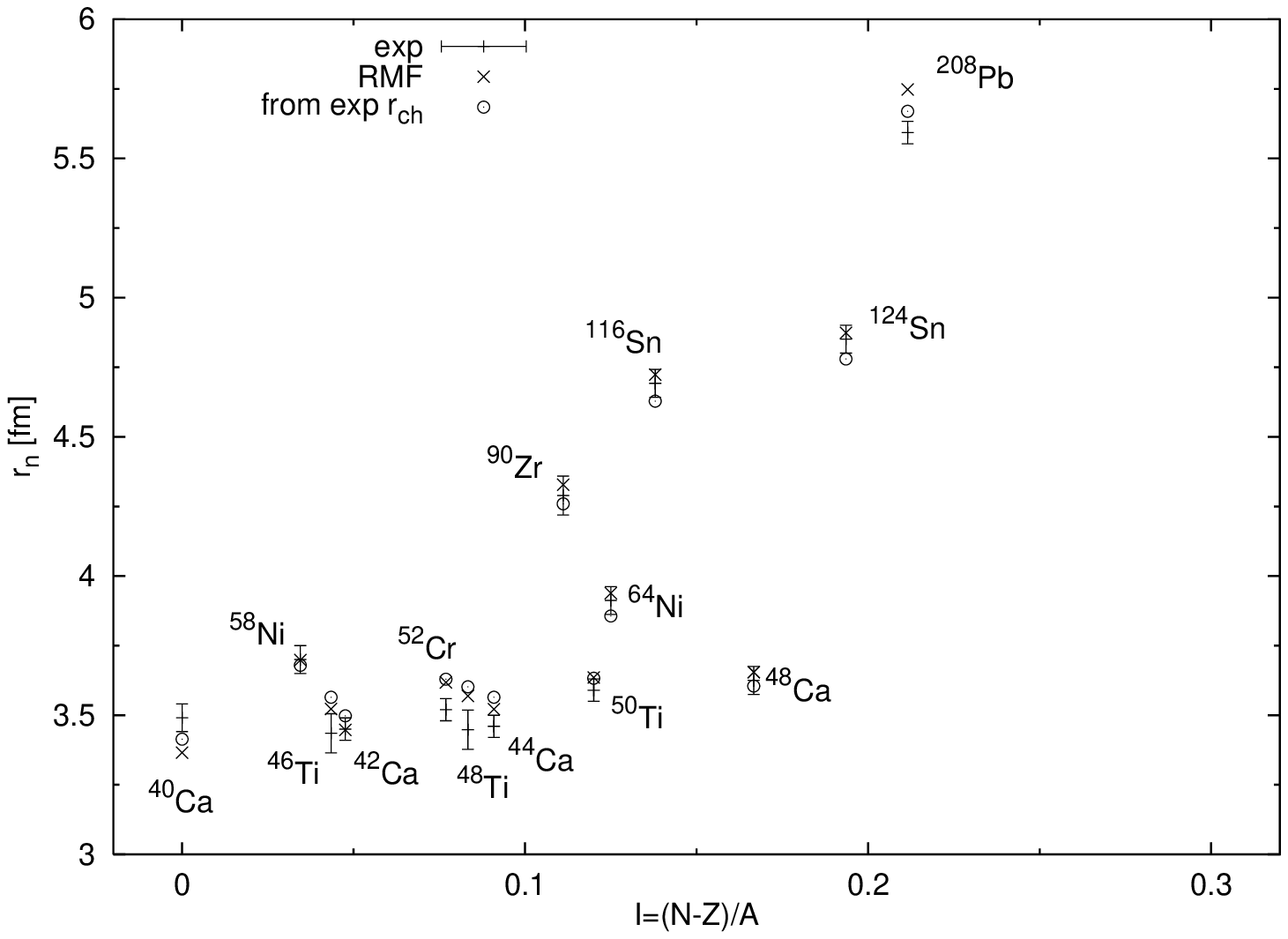}
\caption{ }
\end{figure}

\end{document}